# Modeling Cloud Architectures
# as Interactive Systems

Antonio Navarro Perez and Bernhard Rumpe

Department of Software Engineering,
RWTH Aachen University,
Aachen, Germany

**Abstract.** The development and maintenance of cloud software is complicated by complex but crucial technological requirements that are tightly coupled with each other and with the software's actual business functionality. Consequently, the complexity of design, implementation, deployment, and maintenance activities increases. We present an architecture description language that raises the level of technological abstraction by modeling cloud software as interactive systems. We show how its models correspond to an architecture style that particularly meets the requirements of cloud-based cyber-physical systems. The result provides a basis for an architecture-driven model-based methodology for engineering cloud software.

## 1 Introduction

The development and maintenance of cloud software is complicated by complex but crucial non-functional requirements, for instance: deployment, distribution, scalability, robustness, monitoring, multi-tenancy, and security [1]. These requirements are tightly coupled with each other and with the software's actual business functionality. Consequently, the complexity of design, implementation, deployment, and maintenance activities increases.

Arguably, one of the most complex representatives of cloud-based systems are cyber-physical systems [2] with cloud-based components. In such systems, cloud software interacts with the physical world by monitoring and controlling numerous physical states and processes. Contemporary examples of such systems come from domains such as smart homes [3], smart grids [4], and connected cars [5]. In these systems, cloud software acts as the "brain" of the system by organizing multitudes of different clients, data streams, processes, and calculations. Significantly, all these physical states of affairs are constrained by inherent concurrency, reliability issues, and soft or hard time constraints [6].

In this paper, we introduce an architecture description language (ADL) [7] as the core element of a *model-based* methodology for engineering cloud software. This methodology understands and describes these systems as *interactive systems* [8]. In this context, our ADL describes the *logical software architecture* of such systems. Thereby, it achieves a system representation that abstracts from its mentioned technological requirements. The description of those is, in turn, outsourced to other distinct modeling languages that integrate with the logical software architecture model (e.g., infrastructure models that describe a concrete cloud-based infrastructure setup). Moreover, our ADL supports an architecture style similar to actor-based systems [9] that refines the notion of interactive systems and is tailored to the domain of cloud software.

We hypothesize that (a) interactive systems adequately describe a relevant class of cloud-based systems, that (b) our ADL and its underlying architecture style adequately describes the essence of interactive cloud software, (c) that the resulting overall system model is an adequate representation of the actual system and can be used as a basis for generative techniques, and (d) that our methodology thereby reduces the complexity of the engineering process of such systems.

We begin with a brief overview of related work in Section 2. Subsequently, we introduce the requirements of cloud-based cyber-physical systems in Section 3. We present our ADL and its supported architecture style in Section 4 and discuss it briefly. We conclude with an outlook to future work.

## 2 Related Work

Interactive systems are a special class of reactive systems [10] and commonly described as compositions of independent components (e.g. software, target hardware, external devices) that interact with each other to achieve a common goal. The literature on such systems is extensive. The FOCUS method provides a general formal methodology to describe the architecture of interactive systems through the input/output behavior of components that are statically connected via communication channels. [8, 11] Practical model-based methodologies have been developed that model the architecture of such systems, especially in the domain of real-time embedded systems, for instance, ROOM [12], AADL [13], and MARTE [14]. The ADL MontiArc [15] also models the architecture of such systems according to the FOCUS semantics.

In general, development of cloud-software lacks a set of established standards and tools. [16] In the particular context of cyber-physical systems, Lee et al. discuss a similar need for better specification methods and tools, mentioning modeling and actor-based systems among others. [2, 6] The actor-based paradigm for specifying distributed systems [9, 17, 18] nicely mirrors the essence of interactive systems, addressing in particular the challenge of distribution and scale. Its application in the domain of cloud-based services is recently rising in popularity with frameworks as Scala Akka [19] and Microsoft's Orleans [20].

However, the integration of model-based methodology, actor-based architecture styles, and cloud-based software has not yet been tackled.

## 3 Interactive Cloud Software

We define interactive cloud software as a class of software that drives cloud-based systems. It shares many essential properties and requirements with software in reactive systems, for instance, software in real-time distributed systems. Those properties requirements are:

- *Reactiveness:* the software's main logic is driven by reactions to events from its environment (e.g., sensor signals) or internal events (e.g. interactions between software units). Moreover, it must satisfy given time constraints (e.g., to control an actuator device in time to react to an environmental event).
- *Statefulness:* the software maintains a conversational state with the environment it interacts with (e.g., the current operational state of a device).
- *Concurrency:* events in the real world occur simultaneously and at any given time. Consequently, the software has to have the capability to deal with these events in time of their occurrence, that means, to deal with them concurrently.
- *Distribution:* the software communicates with highly distributed clients over networks with uncertain properties. Moreover, the software itself is running on distributed cloud-based infrastructure.
- *Scalability:* to assure the requirements of timely and concurrent reactiveness, the software has to able to scale dynamically to meet load demands.
- *Reliability:* the software controls aspects of the physical world and, hence, has to meet high standards of reliability. Moreover, the quality of large, distributed, heterogeneous systems is difficult to assure. This is especially true for stateful systems as erroneous states might multiply over time in absence of resetting or cleaning countermeasures.

## 4   Architecture Modeling

The cloudADL (cloud architecture description language) is a textual *Components & Connectors* ADL [7] that describes architectures through hierarchically decomposed, interconnected components. It is an extension of MontiArc [15] and developed with the language workbench MontiCore [21]. It supports an architecture style tailored to systems as described in Section 3.

Architecture models in the cloudADL describe the *logical architecture* of the system. The logical architecture describes the essential structure of the system and the abstract, generalizable properties of its elements. However, it abstracts from specific, non-generalizable implementation details. Thus, cloudADL models do not explicitly define or constraint the technological implementation of modeled concepts or the granularity of how modeled elements are mapped to their counterparts in the real system. For instance, individual components (as introduced in the following section) may represent small software modules (e.g., Java classes), applications (e.g. web applications), or an entire integrated systems. Regardless of that, a component's realization (which in the following is referred to as a runtime component instance) always reflects the same basic, abstract properties expressed by the architecture model.

Figure 1 shows the architecture of a simple cloud service that receives incoming data streams from internet-connected physical sensor devices and stores them to a database. The Figure shows a graphical representations as well as the textual syntax. We will describe this example in the following.

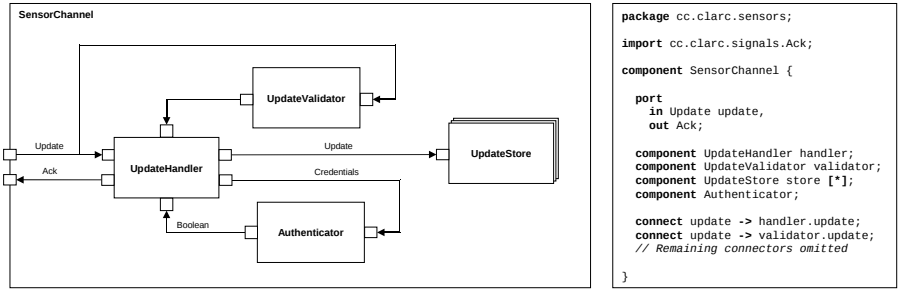

**Fig. 1.** The architecture of a cloud service that receives and stores data streams from sensors.

### 4.1 Concepts and Semantics

In this Section we present the structural elements of the cloudADL. We also informally describe their semantics. As an extension of MontiArc, it shares with it the concepts of component types, subcomponents, and connectors, as well as the underlying formalism from FOCUS [11]. Its most important additions to these are replicating subcomponents, and contexts.

**Components and Interfaces** *Components* are the central building blocks of architectures and represent distinct and independent units of functionality. They interact with each other by *asynchronous [22] message passing* over directed *message channels* that connect an outgoing and an incoming *port* of two components. Ports are typed endpoints for channels and can either receive or send typed messages. Taken together they form the import and export interface of a component.

Channels are associated with *streams*. The notation of streams depicts sequences of messages that have been sent over these channels as seen from a given point in time. Streams, hence, represent the histories of interactions a component has conducted with other components over its incoming and outgoing ports. Their entirety constitutes the visible *behavior* of a component.

The behavior of a component is the specification that defines how the component reacts to incoming messages. It can react by changing an internal state, by sending outgoing messages, by raising an error, or by doing nothing at all. A component's behavior can be defined through additional models (e.g., code of a traditional programming language, or through automaton models) or through *decomposition*. A decomposed component is internally divided into a network of *subcomponents* that defines its behavior. Subcomponents are connected via directed *connectors* between two ports, representing the message channels between them.

In Figure 1 `SensorChannel` is a component with an incoming port that receives messages of type `Update` and an outgoing port that sends messages of type `Ack`. It receives updates from sensor devices and responds with positive or negative acknowledgements. This behavior of `SensorChannel` is defined through its decomposition into four subcomponents that are interconnected with each other and with their parent component's ports. `UpdateHandler` takes incoming updates and organizes their processing by the other components. It passes the update's credentials to the `Authenticator` which, in turn, sends back an approval or a rejection. The `UpdateValidator` also receives the update message, checks its payload for validity, and sends the result to the `UpdateValidator`. Finally, the `UpdateStore` takes accepted updates and writes them into a database. Through collaboration, the network of subcomponents implements the functionality of `SensorChannel`.

Components are defined as types, each in its own model. The type definition contains the declaration of the component's ports, its subcomponents, and its connectors. Subcomponent declarations are *prototype instances* of component types which, in this case, are defined in other models. Likewise, port declarations also refer to externally defined types from numerous sources (e.g., types defined by Java classes, or types from UML/P class diagrams [23]).

The use of component types, with the clear distinction between a type's definition and usage, as well as the mechanism of decomposition allows the modeler to compose a concrete architecture by reusing components. This is further supported by the strong encapsulation of a component's behavior. A component type only exports its interface to other component. Its internal structure, be it a network of subcomponents or any other behavior specification, is kept private. Thus, component realizations can be replaced as long as the component's interface does not change.

**Replication**  Subcomponents can be tagged as *replicating* subcomponents. By default, a subcomponent correlates to exactly one runtime instance of that subcomponent's realization in the actual system. Replicating subcomponents, however, may have a variable number of runtime instances. In this way, the architecture model accounts for parts of the software that can dynamically scale at runtime according to dynamic parameters (e.g., the system's current load) by increasing or decreasing the number of those parts that correlate to a replicating subcomponent. The actual replication strategy is left unspecified.

In Figure 1, the `UpdateStore` subcomponent is tagged as replicating. As I/O operations on a database potentially block a system, it can dynamically replicate itself, for instance, based on the frequency of incoming messages with payload to be stored. Note that the component's replicability remains an abstract property and does not imply a particular technical replication mechanism from instantiating new threads to launching new virtual machines.

Message channels attached to replicating components guarantee that every message is received by exactly one replica. Consequently, the sending endpoint of that channel is agnostic about the eventual replicability of the receiver as, in any case, a message is received by one component instance. However, outgoing ports may be tagged as *replicating ports*. Such ports reflect the number of receivers on the other end of the channel and allow for the explicit selection of receiving component instances. Thus, a component's behavior specification is able to implement alternate communication patterns like broadcasting.

**Contexts** The introduction of replicating components leaves the semantics of message channels underspecified. Channels that connect one or two replicating subcomponents do not give a concrete mechanism for selecting the receiving replica of individual messages. This mechanism can be based on many viable strategies (e.g., round robin), some of which cannot be specified on the level of the logical software architecture (e.g., the selection of the first idle component).

In many scenarios, however, the selection of a particular replica is crucial. For instance, cyclic paths of components and channels might represent a causal chain of interactions in which messages originating at the end of the chain are required to be received by the same component instance that caused the chain in the first place. To give an example, in a web system a set of replicas might be associated to a user session and, thus, require mutually exchanged messages to be only delivered to other replicas of the same set.

Consider the example in Figure 2. A message being received by component A might cause a chain of logically related messages being sent from A to B to C, and from C back to A. Depending on the desired semantic of the message channel between C and A, messages from C might be required to be received by that concrete instance of A that caused the chain of messages in the first place. Without further information, the selection of the concrete replica of A to receive an individual message is ambiguous.

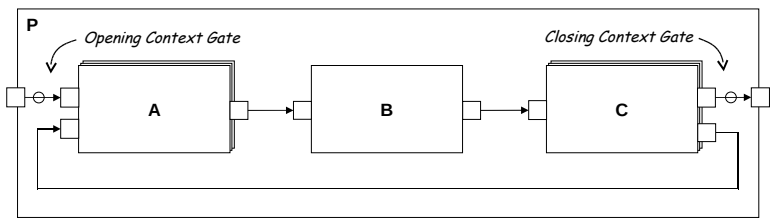

**Fig. 2.** A set of subcomponents with receiver selection ambiguities.

*Contexts* are a concept to address these ambiguities. A context is a class of *context tokens* that are valid in the scope of a decomposed component. It is defined by a name and a set of *opening* and *closing context gates*. A message being sent through an opening context gate is *contextualized* by having a new token of that context assigned to it. Conversely, a message being sent over a closing context gate has its respective token removed. Whenever a contextualized message is received by a component instance, this instance is as well contextualized with the message's context token. Messages and component instances can be contextualized in multiple contexts at the same time.

Contexts can be seen as a generalized mechanism to maintain *conversational states*. A conversational state is a shared state between two or more communicating entities that is defined by the history of communication between them. In our case the entities are components and their communication history is the set of messages they exchanged prior to a certain point in time.

Figure 2 shows two context gates attached to the "first" and "last" connector in P. In this example, all messages received by P are assigned to a token of that context. All messages sent by P have the respective context token removed.

Components that process messages and produce new messages in response are supposed to *handover* context tokens from incoming messages to those outgoing messages that are logically caused by them. In other words, every outgoing message may be the logical result of a set of previously received incoming messages. If this is the case, the outgoing message is contextualized to all contexts of all the incoming messages that logically caused it to be sent.

Usually, the handover can be defined by a simple handover function. For instance, if the behavior specification of a component defines a distinct sequence of actions that are executed for every incoming message, all outgoing messages that are caused by those actions can automatically be contextualized with the context of that incoming message.

To summarize, contexts can be understood as a generalization of the session and session token mechanism in most web systems with client interaction.

## 4.2 Architecture Style

The concepts of the cloudADL are reflected in the actual system in the form of a particular *architecture style*. In essence, the hierarchy of components in the model corresponds to a hierarchy of system parts—the runtime components—, each of which is associated to a component in the model. Similarly, channels between components correspond to interfaces between those system parts. Moreover, those system parts and interfaces all share common properties that are all results of the concepts of the cloudADL.

**Distribution By Default** Runtime components are executed in their own thread of control. Thus, they cannot mutually block or crash each other. They influence each other only through asynchronous message passing over explicit message channels with FIFO-buffers at the receiving endpoint. Their internals are encapsulated and cannot be accessed directly.

Moreover, the behavior of runtime components is, with the exception of replicating ports, agnostic about the interaction partners of its component and about the technical realization of the component's adjacent message channels. A component's behavior implementation receives and sends messages via ports but cannot reflect on what is beyond its ports.

Consequently, the interaction semantics are, by default, of a distributed nature. Functionally, it does not make a difference whether two interacting components are actually distributed or whether they are executed in the same runtime environment. Message channels can be implemented in numerous ways, from local inter-thread communication to dedicated message queue servers. In any case, interactions are always implemented as discrete, asynchronously exchanged typed messages on channels connecting two runtime components. Moreover, the preservation of the order of messages on an individual message channel is guaranteed.

As a result, the architecture style makes physical distribution and networking transparent. The system can be functionally understood without taking its distribution into consideration. Moreover, the physical distribution can be changed without influencing the functional properties of the system.

**Supervisor Hierarchy** The hierarchy of components in the model translates to a hierarchy of *supervision* in the runtime system. Decomposed components represent management logic, so called *supervisors*. Supervisors monitor and manage the component instances associated to their component's immediate subcomponents. These, in turn, can again be other supervisors or functional components. Supervisors manage, in particular, the replication of runtime components associated to replicating subcomponents, and error handling.

The supervisor's error handling aims to deal with errors as locally as possible. To this end, component instances can implement strategies to deal with internal errors. If they fail, they escalate the error up the hierarchy to their supervisor which, again, implements a strategy to deal with the error. This pattern applies until the error is resolved (e.g., by resetting a component instance) or the root of the hierarchy is reached.

**Discussion** This architecture style bears resemblance to actor-based systems. [9] However, there are differences. For instance, actors can dynamically instantiate other actors, whereas the general architecture of component instances is, apart from replicating component instances, static. Moreover, actors have one queue for incoming messages, whereas component instances have as many typed queues as they have incoming ports.

Apart from that, our architecture style reflects the properties and requirements mentioned in Section 3 in a similar way. Distribution is addressed through distribution transparency. Statefulness is implemented by the encapsulated state of components. Concurrency is inherent to the system due to each component's independent thread of control. Scalability and reliability are given through supervision. Finally, reactiveness is achieved as a combination of a concurrent, non-blocking, fault-tolerant, and scalable software architecture.

## 5 Conclusion and Future Work

We have presented an architecture description language that describes the logical software architecture of cloud software in terms of interactive systems. This language is a Components & Connectors language that, most importantly, introduces replication, contexts, and service interfaces as cloud-specific architectural concepts. Furthermore, we have shown how this description maps to the architecture style of distributed, hierarchically supervised components with asynchronous, buffered message passing. We have argued that this architecture style meets the requirements of the class of cloud software that drives cloud-based cyber-physical systems.

We are employing this language in the context of a broader model-based, generative methodology for engineering cloud software named *CloudArc*. This methodology is developed in the context of the SensorCloud project [24] funded by the German Federal Ministry of Economics and Technology, as well as in the context of other projects.

This methodology centers around a logical architecture model and augments it with additional modeling languages specific to other aspects of the overall system. It employs generative techniques in the form of a code generator product line. These code generators synthesize middleware that implements the architecture style described in this paper on various technological infrastructures.

Deployment models are a crucial aspect of the modeling languages in CloudArc. These models describe an infrastructure architecture independent of the logical software architecture and use a mapping model to relate the two. The resulting deployment description is leveraged by the code generators to produce a middleware implementation that reflects the software's target infrastructure. The system's business logic is subsequently implemented through handwritten code.

Test specification models allow for the model-based testing and simulation of the system without the need to deploy it onto production infrastructure. Scenario models and input/output specifications are employed to test the business functionality on a particular variant of middleware that simulates a distributed system but runs locally.

These concepts are currently under development. We plan to evaluate them systematically in the future in the context of our projects.